# AN INTUITIVE DESIGN APPROACH FOR IMPLEMENTING REAL TIME AUDIO EFFECTS

Mayukh Mukhopadhyay[1] and Om Ranjan[2]
[1]Tata Consultancy Services, Kolkata-700091, India
[2]CSIR-CEERI, Pilani, India

*ABSTRACT*

*Audio effect implementation on random musical signal is a basic application of digital signal processors. In this paper, the compatibility features of MATLAB R2008a with Code Composer Studio 3.3 has been exploited to develop Simulink models which when emulated on TMS320C6713 DSK generate real time audio effects. Each design has been done by two different asynchronous scheduling techniques: (i) Idle task Scheduling and (ii) DSP/BIOS task Scheduling. A basic COCOMO analysis has been done for the generated code to justify the industrial viability of this design approach.*

## I. INTRODUCTION

For years musicians have been using different techniques to give their music a unique sound. Some of these techniques were found by serendipity, while others were found after a lot of work and experimentation. While the effects have not changed much in the past few years, the ways in which these effects are produced have.

In the past the effects were implemented with analog technology, or in some cases two tape reels spinning at different speeds [4]. In recent years there has been a movement away from analog signal processing towards digital signal processing of music [4]. This movement has allowed for very precise and easily reproduced effects.

With the advent of real time concepts, digital audio effect implementations have taken a new dimension. Scheduling various processes to meet deadlines has become necessary. In this paper, two different asynchronous schedules: (i) Idle task and (ii) DSP/BIOS task have been adopted to produce the desired audio effects. The various audio effects implemented are: (i) Multi Echo (ii) Natural Reverberation (iii) Stress Generation and (iv) Reverberated Chorus.

The approach is intuitive because the Simulink models are designed without any prior calculation of model parameters. Once any of the models is complete and built, it automatically opens Code Composer Studio and compiles the model. When no compilation error is detected, TMS320C6713 DSK automatically interfaces with the real world [1], [2].

Code generated for periodic tasks, both single- and multitasking, runs out of the context of a timer interrupt. The generated code that represents model blocks for periodic tasks runs periodically, clocked by the periodic interrupt whose period is equal to the base sample time of the model. This execution scheduling scheme is not flexible enough for real time audio effect generation systems that must respond to asynchronous events (random musical signal inputs) in real time. Such systems may need to handle a variety of hardware interrupts in an asynchronous, or aperiodic, fashion.

Target Support Package TC6 (MATLAB R2008a) software facilitates modeling and automatically generating code for asynchronous systems using various task blocks. Among those blocks, the C6000 Idle Task block specifies one or more functions to execute as background tasks in the code generated for the model. The functions are created from the function-call subsystems to which the Idle Task block is connected. Another block, the DSP/BIOS Task blocks spawn free-running tasks as separate DSP/BIOS threads. The spawned task runs the function-call subsystem connected to its output [3].

The paper has been organized as follows: Section II talks about the theory behind various audio effects that are implemented. Section III briefly describes the real time schedules pertaining to the paper and, Section IV elaborates the analysis behind real time stress generation audio effect implementation and illustrates all other designs. The COCOMO analysis has been done in section V, and finally conclusions are drawn in section VI.

## II. THEORY OF GENERATED AUDIO EFFECTS

The audio effects implemented for the project are time domain operations with delay as the basic building block.

### 2.1. Multi Echo

To generate a fixed number of multi echoes spaced R sampling periods apart with exponentially decaying amplitudes, one can use a finite impulse response filter with transfer function of the form

$$H(z) = 1 + \alpha z^{-R} + \alpha^2 z^{-2R} + ... + \alpha^{N-1} z^{-(N-1)R} = \frac{1-\alpha^N z^{-NR}}{1-\alpha z^{-R}} \quad (1)$$

where in (1), α is a constant value and N is an integer. An infinite number of echoes spaced R sampling periods apart can be created by an infinite impulse response filter with transfer function of the form

$$H(z) = 1 + \alpha z^{-R} + \alpha^2 z^{-2R} + \alpha^3 z^{-3R} + ...$$
$$= \frac{1}{1-\alpha z^{-R}}, |\alpha| < 1 \quad (2)$$

### 2.2. Natural Reverberation

In a closed space, such as a concert hall, the sound reaching the listener consists of several components: direct sound, early reflection and reverberation.

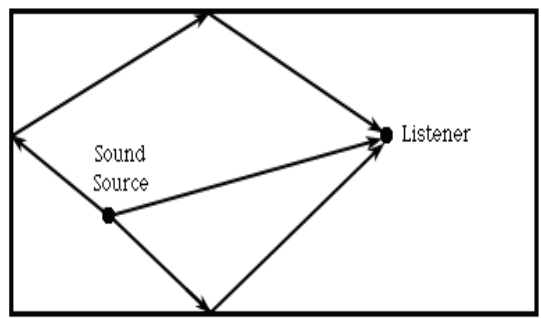

Figure 1. Sound Path [6]

The reverberation is composed of densely packed echoes [4]. It has been observed that approximately 1000 echoes per second are necessary to create a reverberation that sounds free of flutter [5].

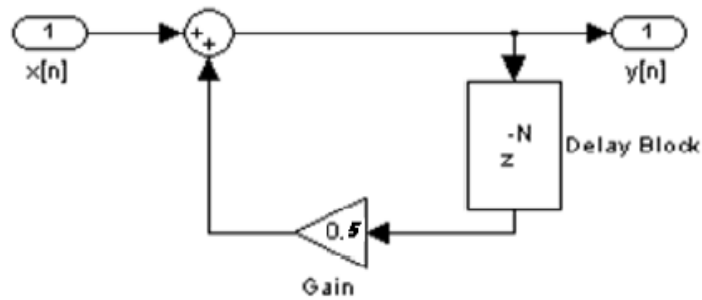

Figure 2. IIR Comb Filter

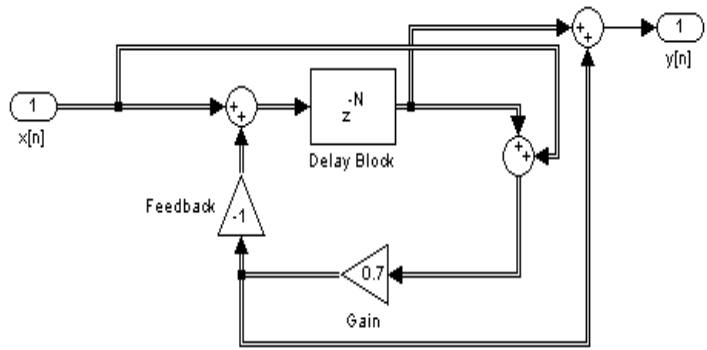

Figure 3. All pass reverberator

The IIR comb filter of Figure 2 and the allpass reverberator of Figure 3 are basic units that are suitably interconnected to develop a natural-sounding reverberation. A proposed scheme of interconnection composed of a parallel connection of IIR echo generators cascaded with two allpass reverberators by [5] has been developed for the desired audio effect.

### 2.3. Stress Generation

This experimental audio effect has been developed to simulate the psychological condition of a stressed individual in cinematography. The effect is generated by cascading delay blocks to the input and the output of an allpass reverberator.

### 2.4. Reverberated Chorus

A reverberated chorus effect is a condition where the listener perceives a single musical note as a choir performance in a concert hall. This effect is generated by summing the outputs of natural reverberation and chorus filter designed by [2]. The occurrence of natural delay due to the asynchronization of different delay blocks has been minimized.

## III. REAL TIME SCHEDULING TECHNIQUES

The purpose of real time computing is to execute, by the appropriate deadlines, its critical control tasks. Each task consumes resources (processor time, memory, and input data) and puts out one or more results. The deadline is the time by which a task must complete its execution [7].

Scheduling refers to the planned allocation of resources for a task in order to produce results within its deadline. The schedule may be preemptive or non-preemtive. A schedule is preemptive if tasks can be interrupted by other tasks (and then resumed). By contrast, once a task is begun in a nonpreemtive schedule, it must run to completion or until it get blocked over a resource.

A task that is currently holding an unsharable resource is said to be in the critical section associated with the resource.

Implementation of binary semaphores is a way to guard the critical section ensuring an exclusive access. Binary semaphores are like locks. When the semaphore is locked (set to 1), it indicates that there is a task currently in the critical section. When a task seeks to enter a critical section, it checks if the corresponding semaphore is locked. If it is, the task is stopped and cannot proceed further until that semaphore is unlocked. If it is not, the task locks the semaphore and enters the critical section. When a task exits the critical section, it unlocks the corresponding semaphore [7].

The idle task block of Target Support Package TC6 creates a free running task which executes the downstream subsystem function. All tasks executed by this block are of the lowest priority, lower than that of the base rate task and preemptively/nonpreemptively scheduled.

The DSP/BIOS task block of Target Support Package TC6/DSP/BIOS Library executes the downstream subsystem function as a separate critical thread of the highest priority. A semaphore is used to enable the DSP/BIOS task execution.

Each audio effect is implemented and executed both preemptively and critically as discussed in the later sections.

## IV. STRESS GENERATOR IMPLEMENTATION AND OTHER AUDIO EFFECT DESIGNS

### 4.1. Mathematical Analysis of Stress Generator

Psychological stress has been defined as a psychological state of tension that is a response to a perceived threat, accompanied by a specific emotion such as fear, anxiety, or anger [8]. During this emotional condition, certain sentences strike the human mind repeatedly after a certain interval of time.

In this paper, a transfer function has been designed to realize an audio effect which simulates this human nature.

If $\alpha$ is an imaginary $n^{th}$ root of unity then:

$$\alpha^{n+1} = \alpha^n . \alpha = 1 . \alpha = \alpha \quad (3)$$

Let the three roots of unity be 1(real), $\alpha$ (complex) and $\alpha^*$ (complex conjugate). Then the transfer function needed to be established is:

$$h(n) = [A + B\alpha^n + C(\alpha^*)^n].u(n) \quad (4)$$

where in (4), A, B and C are constants, and $u(n)$ is an unit step sequence. The equation after applying z- transform on both sides of (4) is:

$$H(z) = \frac{A}{1 - z^{-1}} + \frac{B}{1 - \alpha z^{-1}} + \frac{C}{1 - \alpha^* z^{-1}} \quad (5)$$

Again,

$$\alpha . \alpha^* = 1 \quad (6)$$

Hence,

$$H(z) = \frac{A}{1-z^{-1}} + \frac{B\alpha^*}{\alpha^*-z^{-1}} + \frac{C\alpha}{\alpha-z^{-1}}$$

$$= \frac{A(\alpha^*-z^{-1})(\alpha-z^{-1}) + B\alpha^*(1-z^{-1})(\alpha-z^{-1}) + C\alpha(1-z^{-1})(\alpha^*-z^{-1})}{(1-z^{-1})(\alpha^*-z^{-1})(\alpha-z^{-1})}$$

$$= \frac{A[1-(\alpha+\alpha^*)z^{-1}+z^{-2}] + B[1-(1+\alpha^*)z^{-1}+\alpha^* z^{-2}] + C[1-(1+\alpha)z^{-1}+\alpha z^{-2}]}{(1-z^{-1})[1-(\alpha+\alpha^*)z^{-1}+z^{-2}]} \quad (7)$$

For the transfer function design, $\alpha$ is taken as the $8^{th}$ root of unity. Hence,

$$\alpha = -\frac{1}{\sqrt{2}}(1+j) = e^{j\frac{5\pi}{4}} \quad (8)$$

$$\alpha^* = -\frac{1}{\sqrt{2}}(1-j) = e^{-j\frac{5\pi}{4}} \quad (9)$$

$$\alpha^9 = e^{j\frac{45\pi}{4}} = e^{j(10\pi+\frac{5\pi}{4})} = (\cos 10\pi + j\sin 10\pi)e^{j\frac{5\pi}{4}} = (1+j0)e^{j\frac{5\pi}{4}} = e^{j\frac{5\pi}{4}} = \alpha \quad (10)$$

$$\alpha + \alpha^* = -\sqrt{2} \quad (11)$$

Where in (8), (9) and (10), $j = \sqrt{-1}$. The denominator of (7) becomes,

$$(1-z^{-1})[1-(\alpha+\alpha^*)z^{-1} + z^{-1}z^{-2}]$$
$$= (1-z^{-1})[1-\sqrt{2}z^{-1}+z^{-2}]$$
$$= 1 + 0.414z^{-1} - 0.414z^{-2} - z^{-3} \quad (12)$$

And the numerator of (7) reduces to,

$$(A+B+C) - [A(\alpha+\alpha^*) + (B+C) + (B\alpha^* + C\alpha)]z^{-1} + (A+B\alpha^* + C\alpha)z^{-2} \quad (13)$$

The simplest form of (13) can be achieved when coefficient of $z^{-2}$ equals to zero. That is,

$$A + B\alpha^* + C\alpha = 0 \quad (14)$$

The possibility of (14) is in affirmative only when B and C are equal. That is using (11),

$$A + B(\alpha^* + \alpha) = 0$$
$$A - \sqrt{2}B = 0$$
$$A = \sqrt{2}B \quad (15)$$

Using (15) in (13), the numerator again reduces to,

$$(\sqrt{2}+2)B - (-2B + 2B - \sqrt{2}B)z^{-1}$$
$$= \frac{\sqrt{2}B}{\sqrt{2}-1}(1+0.414z^{-1}) = K(1+0.414z^{-1}) \quad (16)$$

where in (16), $K = \frac{\sqrt{2}B}{\sqrt{2}-1}$. Hence the final transfer function for stress generator, using (7), reduces to,

$$H(z) = \frac{K(1+0.414z^{-1})}{(1+0.414z^{-1} - 0.414z^{-2} - z^{-3})} \quad (17)$$

The Simulink model implementation of (17) has been shown in Figure 4.

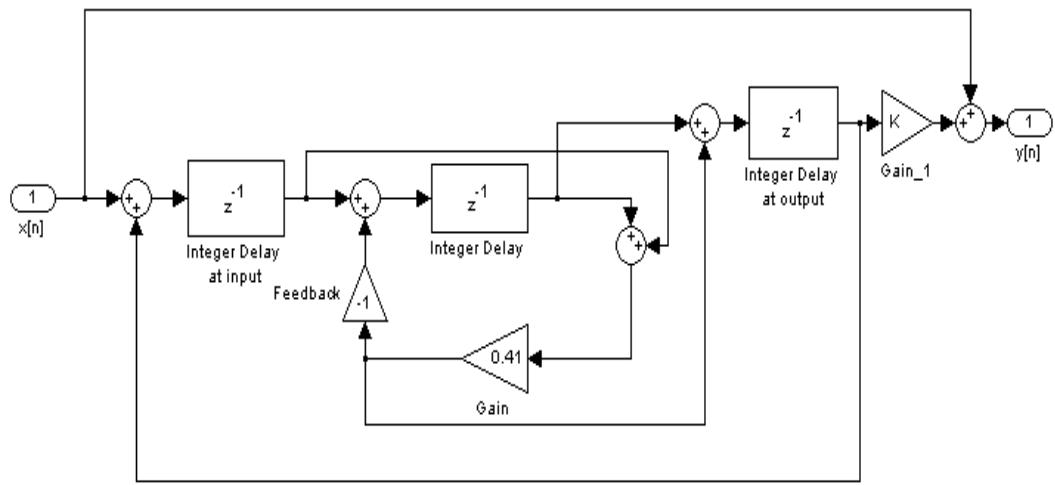

Figure 4. Simulink model implementation of $H(z)$

## 4.2. Real time implementation of Stress Generator

A Simulink model of real time stress generator using (17) has been designed as depicted in the Figure 5.

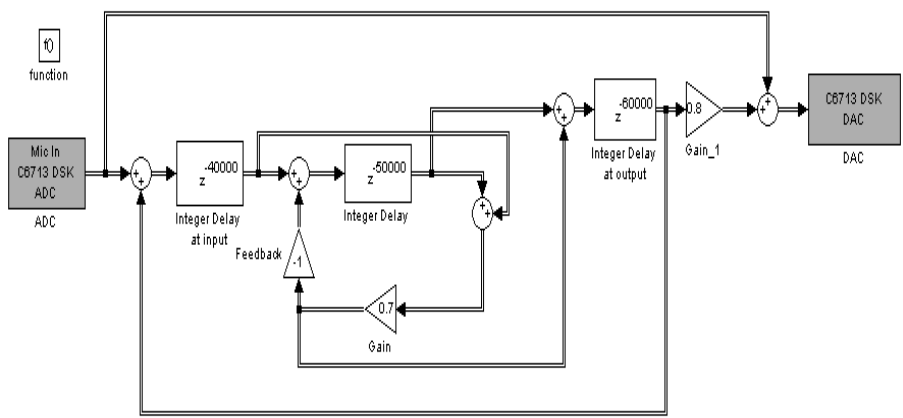

Figure 5. Simulink model of Real Time stress generator

Figure 5 has been implemented in two different ways as depicted in Figure 6 and Figure 7.

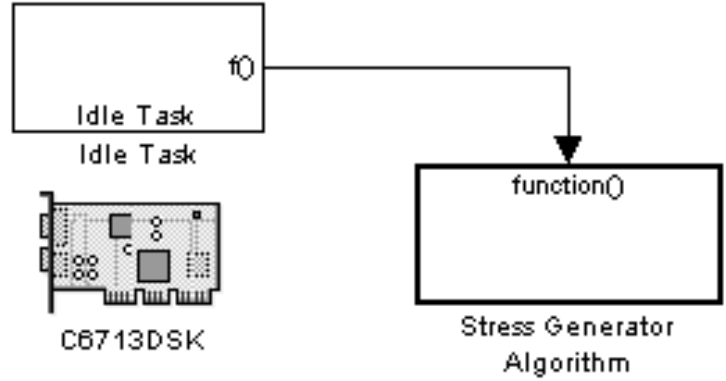

Figure 6: Preemptive Scheduling of Real Time Stress Generator

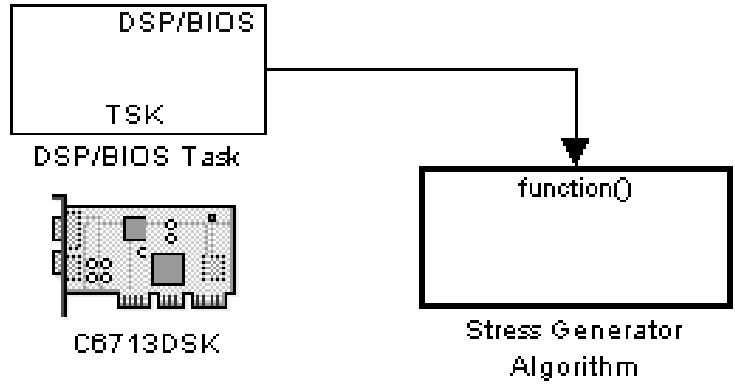

Figure 7: Critical Scheduling of Real Time Stress Generator

The mathematical analysis of the model depicted in Figure 5 has been summarized in (18) to (24).

$$H(z) = \frac{0.8(1+0.7z^{-50000})}{(1+0.7z^{-50000} - 0.7z^{-100000} - z^{-150000})} \quad (18)$$

$$\alpha = -0.85 + j0.527 \quad (19)$$

$$\alpha^* = -0.85 - j0.527 \quad (20)$$

$$h(n) = [A + B(\alpha)^{\frac{n}{50000}} + C(\alpha^*)^{\frac{n}{50000}}].u(\frac{n}{50000}) \quad (21)$$

$$A = 1.36 \quad (22)$$

$$B = -j[\frac{0.759(\alpha + 0.7)}{(1-\alpha^*)}] \quad (23)$$

$$C = j[\frac{0.759(\alpha^* + 0.7)}{(1-\alpha)}] \quad (24)$$

## 4.3. Outline of other real time audio effect designs

On the lines of stress generator, other designs can be similarly implemented.

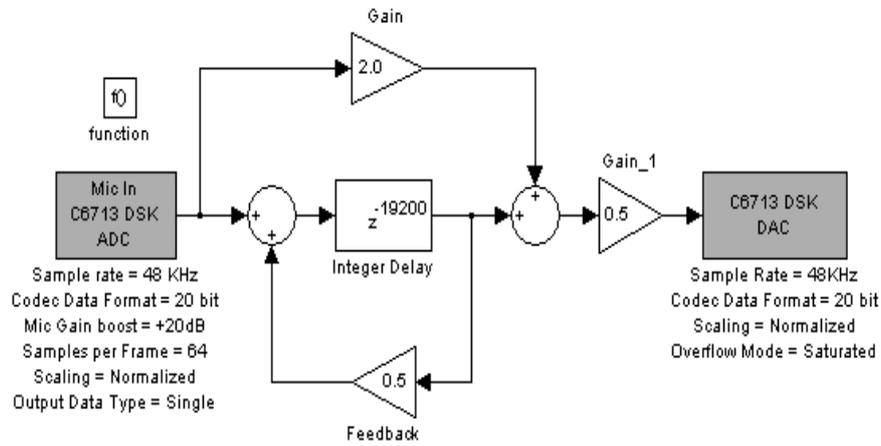

Figure 8: Simulink model of Multi Echo

Figure 8 has been a work around model of Figure 2. This modification has been done to overcome the algebraic loop problem pertaining to generation of C language codes from Simulink models.

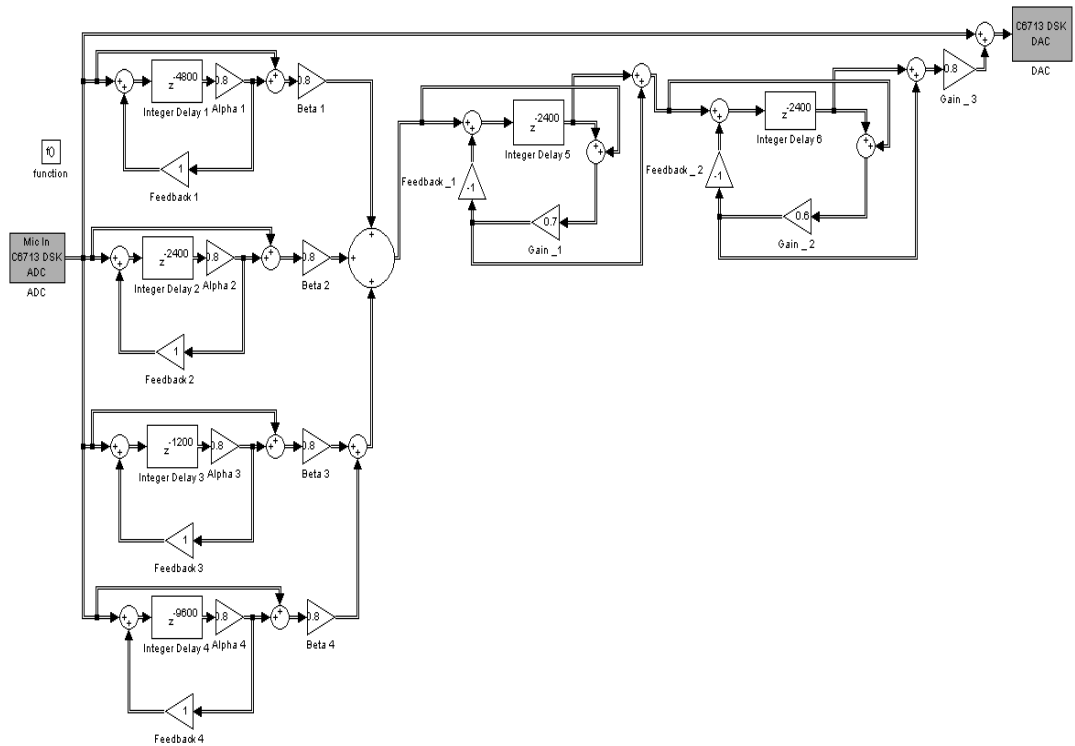

Figure 9. Simulink model of Natural Reverberation

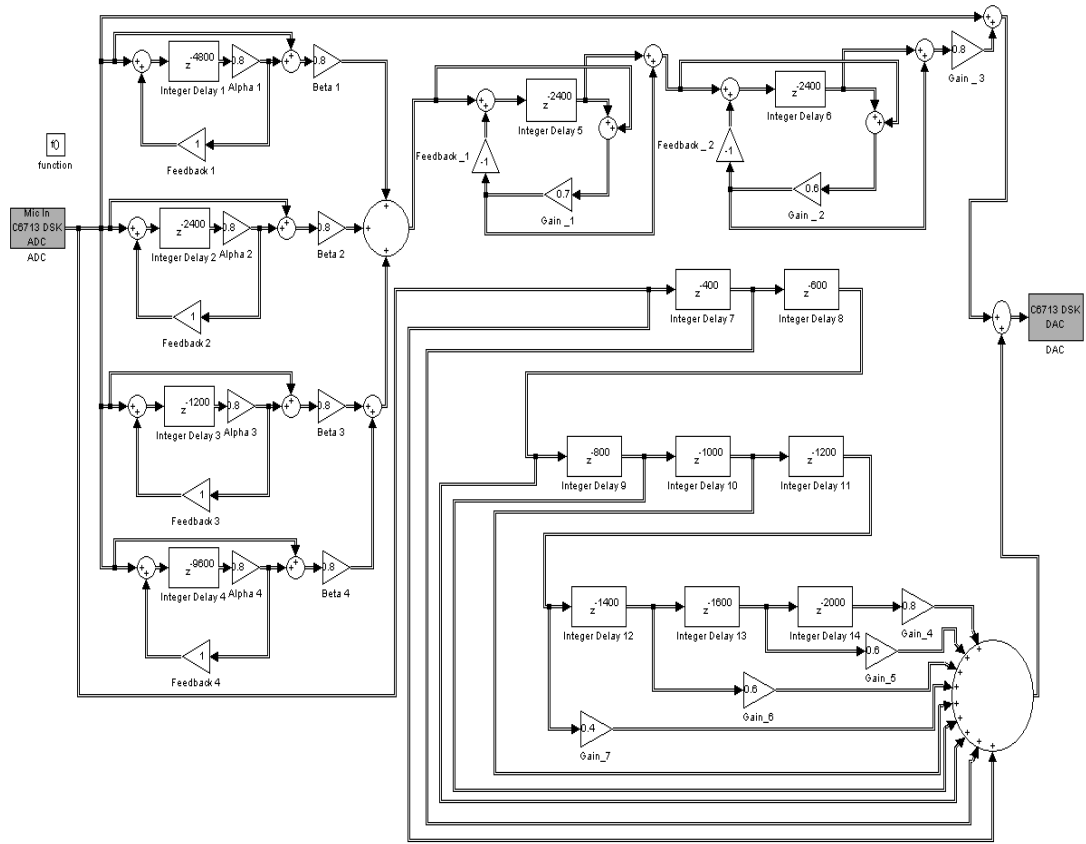

Figure 10. Simulink model of Reverberated Chorus

For Figure 8, Figure 9 and Figure 10 the integer delay and gain values has been intuitively chosen for comfortable audio outputs. The parameters of ADC (analog to digital converter) and DAC (digital to analog converter) used in Figure 8 are same for all other Simulink models.

## V.  BASIC COCOMO ANALYSIS

Constructive Cost Model (COCOMO), developed by [9] has been used to estimate the cost of the project.

According to [9] any software development project can be classified into one of the following three categories based on the development complexity: organic, semidetached and embedded. A development project is considered to be organic type, if the project deals with a well understood application program, the size of the development team members are experienced in developing similar types of projects. When the development team consists of a mixture of experienced and inexperienced staff, the project is identified as semidetached. A development project is considered to be embedded type, if the software being developed is strongly coupled to complex hardware, or if stringent regulations on operational procedures exist.

The flowchart for COCOMO analysis has been summarized in Figure 11 where a, A, b and B are project specific heuristic constants and KLOC is the estimated size of the software product expressed in kilo lines of uncommented code.

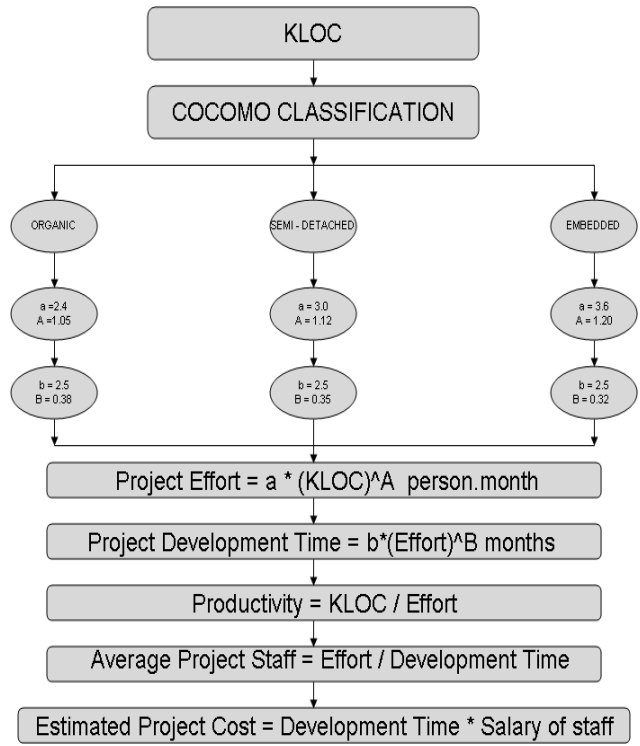

Figure 11: Flowchart for COCOMO Analysis

Table 1, Table 2, Table 3 and Table 4 show the input parameters, standard output parameters, actual output parameters and conclusive advantage of our project using COCOMO analysis respectively.

Table 1. Input parameters for COCOMO analysis

| Project size | 16000 LOC |
|---|---|
| KLOC | 16 |
| Project Type | Embedded |
| Heuristic Constants | a= 3.6<br>A= 1.20<br>b= 2.5<br>B= 0.32 |

Table 2. Standard output parameters

| Project effort | 100.287 person.month |
|---|---|
| Development time | 10.922 months |
| Productivity | 159.54 LOC/person.month |
| Average Project Staff | 9.18 project staff |
| Estimated Project Cost | Rs. (11*18000) = Rs. 198,000 |

Table 3. Actual output parameters using our design approach

| Project effort | 8 person.month |
|---|---|
| Development time | 2 months |
| Productivity | 2000 LOC/person.month |
| Average Project Staff | 4 project staff |
| Estimated Project Cost | Rs. (2*18000) = Rs. 36,000 |

Average industry level salary has been arbitrarily chosen as Rs. 18000 per man.month for project cost estimation.

Table 4. Conclusive advantage of our design approach

| Project effort | Reduced | By 10 times |
|---|---|---|
| Development time | Reduced | By 5.5 times |
| Average Project Staff | Reduced | By 1.8 times |
| Productivity | Increased | By 10.02 times |
| Estimated Project Cost | Reduced | By 5.5 times |

## VI. CONCLUSION

Two different design approaches have been adopted for implementing each of the four audio effect algorithms. The clarity of same audio effect is better when the DSP/BIOS task scheduling was employed with respect to idle task scheduling. Hence it has been concluded that. TMS320C6713 digital signal processor gives finer output when the audio effect algorithms are implemented as a critical control task using semaphores as locks rather than using an idle, preemptive and background task of lowest priority. Further using table IV it can be concluded that our design approach is industrially advantageous with respect to development of same the audio effect algorithm by coding from scratch. Future work can involve controlling the time duration of stress generated signal by manipulation of control function and study its effect on synaptic impulses in living organism.

### ACKNOWLEDGEMENTS

The authors would like to thank Techno India College of Technology, New Town, Kolkata for providing the necessary hardware equipments to verify the simulations in their DSP lab. The authors are grateful to Ardhendu Shekhar Biswas for guidance and valuable technical inputs. The authors also appreciate necessary motivation from Sumitava Karmakar, Md Habib and Anirban Kanungoe.

### REFERENCES

[1] Gannot and Arvin: A Simulink and Texas Instruments C6713 based Signal Processing laboratory, 2006.
[2] S K Hasnain , Implementation of DSP Real Time Concepts Using Code Composer Studio 3.1, TI DSK TMS320C6713 and DSP Simulink Blocksets, Nighat Jamil Pakistan Navy Engineering College (NUST) Karachi, Pakistan.
[3] MathWorks, Schedulers and Timing: Targeting C6000 DSP Hardware (Target Support Package TC6) MATLAB 7.6.0.324(R2008a) February 10, 2008.
[4] Sanjit K. Moitra, Digital Signal Processing- A Computer Based Approach, Chapter 14, Section 14.5, pp. 821- 827, Tata McGraw Hill, 2000.

[5] Schroeder, M.R., "Natural Sounding Artificial Reverberation," Journal of the Audio Engineering Society 10(3):219-223,1962.
[6] Vikas Sahdev, 'Sound Processing', Unpublished Paper.
[7] C.M. Krishna, Kang G. Shin, Real Time Systems, pp. 40- 41, 67, McGraw-Hill International Editions, 1997.
[8] Jerome J. Congleton, William A. Jones, Samuel G. Shiflett, Kevin P. Mcsweeney, R. Dale Huchingson, An Evaluation of Voice Stress Analysis Techniques in a Simulated AWACS Environment, pp. 61-69, International Journal of Speech Technology,1997.
[9] B.W. Boehm. Software Engineering Economics. IEEE Transactions on Software Engineering, 10(1):135-152, Jan. 1984.

**Authors**

Mayukh Mukhopadhyay

The Author is an alumnus of ECE Department, Techno India College of Technology, Rajarhat, Kolkata. He has more than 3.5 years of experience in the area of Business Intelligence and Reporting using Oracle. Currently he is working as an ETL developer for British Telecommunications Retail Team in Enterprise Information Platform and pursuing Jadavpur University-TCS collaborative Masters of Engineering in Software Engineering.

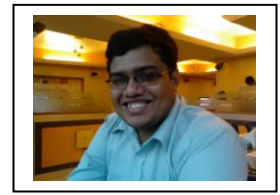

Om Ranjan

The Author is an alumnus of ECE Department, Techno India College of Technology, Rajarhat, Kolkata. He is currently working as a Scientist in CSIR-CEERI labs, Pilani. He holds an M.Tech in High Power Microwave Devices and System Engineering from AcSIR and specializes in application of Sheet Beam Gyrotron.

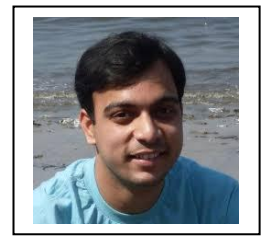